\documentclass[aps,apl,onecolumn,amsfonts,fleqn]{revtex4-1}

\usepackage{graphicx}
\usepackage{amsmath}
\usepackage{amssymb}
\usepackage{relsize}
\usepackage{xcolor}
\usepackage{pifont}
\usepackage[arrowdel]{physics}
\usepackage[exponent-product = \cdot]{siunitx}
\usepackage[colorlinks=true,citecolor=blue,linkcolor=magenta]{hyperref}

\hypersetup{
pdfauthor = {P. Maletinsky},
colorlinks = true, linkcolor = blue, urlcolor=blue, bookmarksnumbered =  true}

\newcommand{\mw}{\rm{MW}}

\begin{document}

\title{Determination of Intrinsic Effective Fields and Microwave Polarizations by High-Resolution Spectroscopy of Single NV Center Spins}

\author{J.~K\"olbl}
\affiliation{Department of Physics, University of Basel, Basel 4056, Switzerland}
\author{M.~Kasperczyk}
\affiliation{Department of Physics, University of Basel, Basel 4056, Switzerland}
\author{B.~B\"urgler}
\affiliation{Department of Physics, University of Basel, Basel 4056, Switzerland}
\author{A.~Barfuss}
\affiliation{Department of Physics, University of Basel, Basel 4056, Switzerland}
\author{P.~Maletinsky}
\email[]{patrick.maletinsky@unibas.ch}
\affiliation{Department of Physics, University of Basel, Basel 4056, Switzerland}

\date{\today}


\begin{abstract}
We present high-resolution optically detected magnetic resonance (ODMR) spectroscopy on single nitrogen-vacancy (NV) center spins in diamond at and around zero magnetic field. The experimentally observed transitions depend sensitively on the interplay between the microwave (MW) probing field and the local intrinsic effective field comprising strain and electric fields, which act on the NV spin. Based on a theoretical model of the magnetic dipole transitions and the MW driving field, we extract both the strength and the direction of the transverse component of the effective field. Our results reveal that for the diamond crystal under study, strain is the dominant contribution to the effective field. Our experiments further yield a method for MW polarization analysis in a tunable, linear basis, which we demonstrate on a single NV spin. Our results are of importance to low-field quantum sensing applications using NV spins and form a relevant addition to the ever-growing toolset of spin-based quantum sensing.
\end{abstract}

\maketitle

\section{Introduction}

The nitrogen-vacancy (NV) center in diamond~\cite{Doherty2013} has long shown promise as an excellent sensor, due to its exceptional sensitivity to external fields. It has demonstrated far-reaching potential in applications ranging from electrometry~\cite{Dolde2011,Dolde2014} and thermometry~\cite{Toyli2013,Neumann2013,Kucsko2013} to precession gyroscopy~\cite{Maclaurin2012,Ledbetter2012,Ajoy2012} and, most notably, magnetometry~\cite{Maze2008,Grinolds2013,Rondin2014}. The NV's success results from a range of useful properties, including its long spin dephasing times~\cite{Balasubramanian2009,Pham2012}, its atomic size, and its ability to be optically initialized and read out~\cite{Gruber1997,Doherty2013}. All of these characteristics contribute to make the NV a highly sensitive system with nanoscale spatial resolution, even at ambient conditions.

The NV spin's impressive sensitivity is, however, also its weakness. Local intrinsic fields arising from lattice strain, paramagnetic impurities, and electric fields induced by surface charges all limit the NV's ability to detect and characterize external signals of interest. 
Techniques operating at low magnetic fields, such as zero-field nuclear magnetic resonance (NMR)~\cite{Weitekamp1983,Thayer1987} and low-field magnetometry~\cite{Zheng2019} are especially vulnerable to these parasitic fields~\cite{Mittiga2018}. Their relevance and future prospects motivate extended studies to precisely characterize the environment surrounding NV spins.

High-resolution spectroscopy offers a set of tools to study the interaction of various local intrinsic and externally applied fields. For example, NV ensembles in type-Ib diamond were probed with microwave (MW) manipulation fields at zero magnetic field to investigate intrinsic effective fields, which represent the combined  effects of strain and electric fields~\cite{Mittiga2018}. This ensemble study revealed that for the particular diamonds under study, the electric field was the dominant contribution to the effective field. A similar technique was introduced to determine the orientation of single NVs~\cite{Doherty2014} by simultaneously applying external electric and magnetic fields. However, a systematic study of the effective fields for individual NV centers in high purity diamond has not been reported so far.

Here, we use high-resolution, low-power optically detected magnetic resonance (ODMR) spectroscopy to characterize in detail the intrinsic effective fields affecting single NVs in high purity, type-IIa diamond. In contrast to the results mentioned above~\cite{Mittiga2018}, we find that in our samples the strain contribution to the effective field dominates over the electric field in low-field ODMR measurements of single NVs. We tentatively assigned this discrepancy to the higher dopant density in the samples of~\cite{Mittiga2018}. Moreover, by applying external magnetic fields and exploiting the magnetic dipole selection rules we directly probe the MW polarization at the NV position. Thus, our method offers a characterization tool for both the intrinsic effective field and the MW manipulation field, paving the way for future sensing applications.

In our experiments, we perform ODMR measurements on a selection of individual NV spins, and thereby extract information about the local environment of each NV defect. Quantitative analysis of our results requires a detailed understanding of the NV spin transition strengths under ODMR driving at low fields. To that end, we first introduce the Hamiltonian describing the hyperfine structure of the NV's ground state in the presence of magnetic, electric, and strain fields. We then present a model to describe how these fields influence the selection rules of the magnetic dipole spin transitions. Based on this model, we calculate the transition magnetic dipole moments and thereby the fluorescence signal we expect to observe in our experiments. By comparing our theory to our experimental data, we are then able to directly characterize the local strain and electric field environment and, in a separate experiment, determine the MW polarization used to drive the spin transitions.

\section{Theoretical background}

\begin{figure*}[htb]
	\centering
		\includegraphics{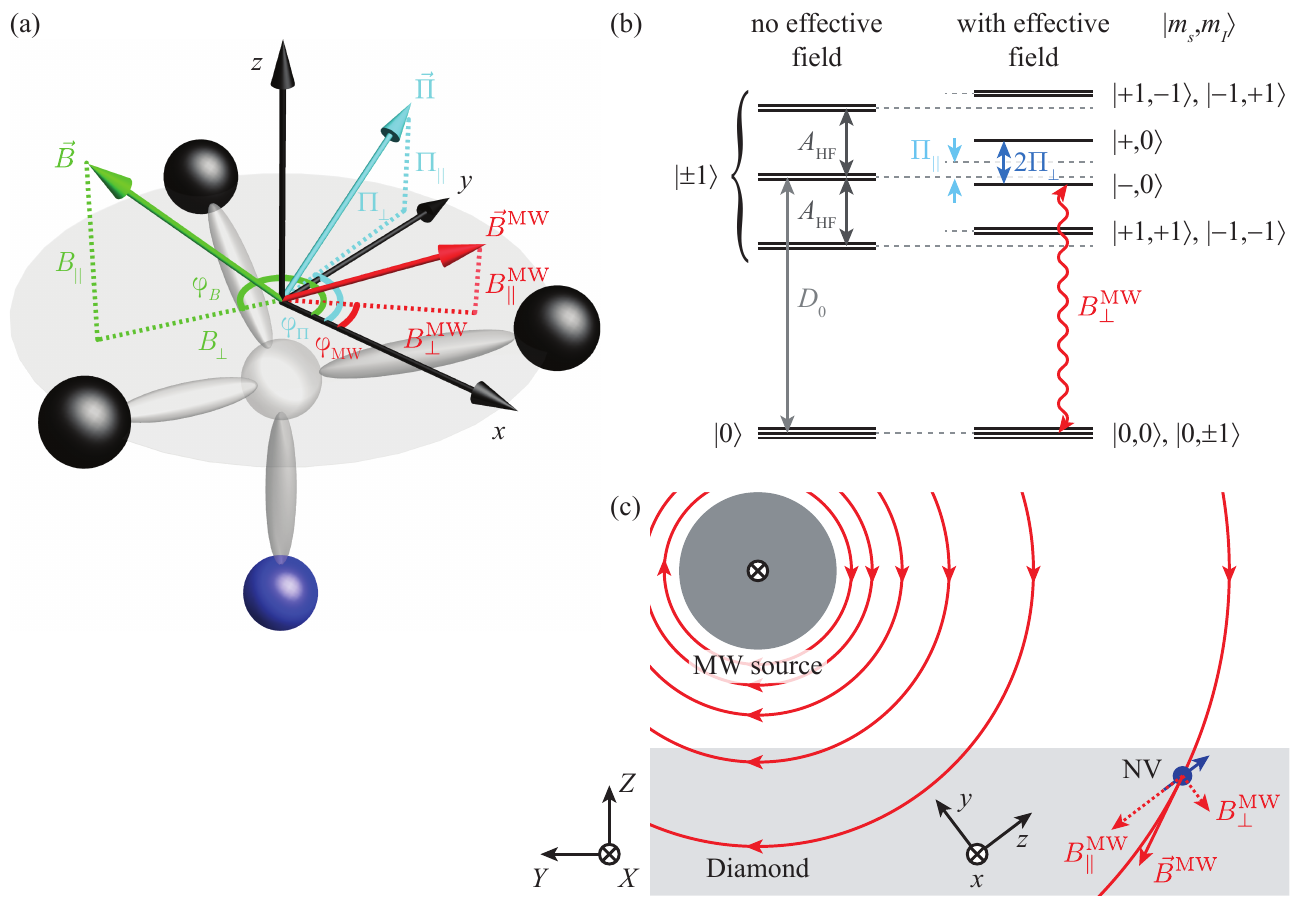}
	\caption{Description of the experimental setup.
	(a) Coordinate system of the NV defect center with the relevant fields interacting with the NV spin: The MW magnetic field (red) for spin manipulation, the effective field comprising strain and electric field (light blue) intrinsic to the diamond sample and the static magnetic field (green) externally applied to determine unknown MW polarizations. All fields are partitioned into components parallel and perpendicular to the NV symmetry axis, where the perpendicular components are further parametrized by the azimuthal angle $\varphi_\alpha$ ($\alpha \in \{ \rm{MW}, \, \Pi, \, \rm{B}  \}$).
	(b) Level diagram of the NV ground state according to Hamiltonian\,\eqref{fml.hamiltonian} without and with the effective field in absence of a magnetic field. The effective field induces a common-mode shift of the $\ket{\pm 1}$ hyperfine levels by $\Pi_\|$ and additionally splits near-degenerate levels which have the same nuclear spin projection (here $m_I = 0$). A MW magnetic field is used to address transitions between the states.
	(c) Schematic of the experimental setup with a diamond sample and a nearby bonding wire for MW delivery. The NV spin under study experiences an approximately linear polarized field which is used to manipulate the NV spin (the situation shown corresponds to the one of NV1).}
	\label{fig1}
\end{figure*}

\subsection{Hamiltonian of the NV center}

The NV center is a paramagnetic lattice defect in diamond comprising a substitutional nitrogen atom adjacent to a lattice vacancy (refer to figure\,\ref{fig1}(a)). The axis joining the nitrogen atom and the vacancy points along diamonds $\expval{111}$ crystal direction and defines the defect's symmetry axis. In the negative charge state we consider here, the NV defect traps an additional electron, resulting in an $S=1$ electronic spin ground state. The corresponding spin eigenstates of the $\hat{S}_z$ operator with respect to the NV axis are the $\ket{m_s}$ spin projection states with eigenvalues $m_s = 0, \pm 1$~\cite{Doherty2013}. Spin-spin interactions lead to a zero-field splitting $D_0$ between the $\ket{0}$ and the degenerate $\ket{\pm 1}$ states. External and intrinsic fields further alter the states' energies. Besides being sensitive to magnetic fields through the Zeeman effect, the $\ket{m_s}$ states are also susceptible to electric and strain fields~\cite{Dolde2011}, caused by spin-orbit and spin-spin mixing of the ground and excited states. Moreover, interactions with the nuclear spin of the $^{14}$N nucleus lead to a hyperfine coupling and additional structure.

Considering all of these interactions leads to the Hamiltonian describing the NV's ground state~\cite{Doherty2013}:
\begin{align}
\mathcal{H}/h = \qty(D_0 + \Pi_z) S_z^2 + \gamma_{\rm{NV}} \vb*{B} \cdot \vb*{S} + A_{\rm{HF}} S_z I_z + \Pi_x \qty(S_y^2-S_x^2) + \Pi_y \qty(S_x S_y + S_y S_x) \ ,
\label{fml.hamiltonian}
\end{align}
where $\vb*{S} = [S_x, S_y, S_z]$ ($\vb*{I} = [I_x, I_y, I_z]$) is the vector of the dimensionless electronic (nuclear) $S=1$ ($I=1$) spin operators of the NV ($^{14}$N) spin and $h$ is the Planck constant. The zero-field splitting is $D_0 \approx \SI{2.87}{GHz}$, and the coupling parameters are the axial hyperfine parameter $A_{\rm{HF}} = \SI{-2.14}{MHz}$ and the NV gyromagnetic ratio $\gamma_{\rm{NV}} = \SI{2.8}{MHz/G}$~\cite{Doherty2013}. The magnetic field $\vb*{B} = [B_x, B_y, B_z]$ and the effective field $\vb*{\Pi} = [\Pi_x, \Pi_y, \Pi_z]$ are given in the coordinate frame ($xyz$) of the NV, where $z$ denotes the NV axis and we choose $y$ to lie in one of the NV symmetry planes (see figure\,\ref{fig1}(a)). The effective field defined as the combined strain and electric field~\cite{Doherty2012} is represented by $\Pi_z = d_\| E_z + \mathcal{M}_z$ and $\Pi_{x,y} = d_\perp E_{x,y} + \mathcal{M}_{x,y}$, where $\vb*{E} = [E_x, E_y, E_z]$ is the electric field and $d_\| = \SI{0.35}{Hz~cm/V}$ and $d_\perp = \SI{17}{Hz~cm/V}$ are the axial and transverse electric field susceptibilities, respectively~\cite{VanOort1990}. The parameters of the spin-strain interaction $\mathcal{M}_{x,y,z}$ weight the components of the strain tensor $\vb*{\varepsilon}$ with the corresponding spin-strain coupling-strength susceptibilities (refer to~\cite{Udvarhelyi2018, Barfuss2019} for more details). Note that Hamiltonian\,\eqref{fml.hamiltonian} neglects the non-axial hyperfine interaction, which is suppressed by the zero-field splitting, the nuclear electric quadrupole interaction, which does not cause a state mixing and therefore does not affect the transition frequencies, and the nuclear Zeeman coupling, as these contributions do not affect the spin states in the parameter regime we consider here.

In absence of magnetic, electric and strain fields, the level structure is dominated by the zero-field splitting, which shifts the $\ket{\pm 1}$ states with respect to $\ket{0}$ by $D_0$ as illustrated in figure\,\ref{fig1}(b). An additional splitting of the $\ket{\pm 1}$ levels is induced by the hyperfine interaction $|A_{\rm HF}|$, leaving three two-fold degenerate hyperfine eigenstates labeled with $\ket{\pm 1,m_I}$, where $m_I = 0,\pm 1$ are the eigenvalues of the $I_z$ nuclear spin operator (see in figure\,\ref{fig1}(b)). Note, that $\ket{0}$ is not affected by the hyperfine interaction, i.e. $\ket{0,m_I}$ is three-fold degenerate.

\subsection{Influence of the effective field}

To quantify the influence of the effective field $\vb*{\Pi}$ on the NV level structure, it is convenient to partition the field vectors into components parallel and perpendicular to the NV axis ($z$-axis), i.e. to write $\vb*{\Pi} = [\Pi_\perp \cos \varphi_\Pi, \Pi_\perp \sin \varphi_\Pi, \Pi_\|]$ (see figure\,\ref{fig1}(a)). Here, $\Pi_\| = \Pi_z$ is the parallel and $\Pi_\perp = (\Pi_x^2 + \Pi_y^2)^{1/2}$ is the perpendicular effective field amplitude. The direction of $\Pi_\perp$ in the transverse $xy$-plane is characterized by the azimuthal angle $\varphi_\Pi$ with respect to the $x$-axis, defined by $\tan \varphi_\Pi = \Pi_y/\Pi_x$. We will treat $\vb*{B}$ the same, with $B_\|$, $B_\perp$ and $\varphi_B$ defined similarly.

According to Hamiltonian\,\eqref{fml.hamiltonian}, the effective field $\vb*{\Pi}$ affects the level structure in two ways: (1) an axial effective field shifts the $\ket{\pm 1}$ states with respect to $\ket{0}$ by an amount $\Pi_\|$; (2) a transverse effective field couples the $\ket{\pm 1}$ states with the same nuclear spin projection. When $B = 0$, the transverse effective field mixes and splits the degenerate $\ket{\pm 1}$ states with $m_I=0$, leading to new spin eigenstates given by
\begin{align}
\begin{split}
\ket{-} &= \qty( e^{i\varphi_\Pi} \ket{+1} + \ket{-1} )/\sqrt{2} \ , \\
\ket{+} &= \qty( e^{i\varphi_\Pi} \ket{+1} - \ket{-1} )/\sqrt{2} \ ,
\label{fml.eigenstates}
\end{split}
\end{align}
where we have omitted the label for the nuclear spin projection $m_I = 0$ for clarity. The corresponding eigen\-energies are $E_{\pm} = D_0 + \Pi_\| \pm \Pi_\perp$ as shown in figure\,\ref{fig1}(b). Thus, besides the level shift of $\Pi_\|$ experienced by all nuclear spin projections, the coupled states $\ket{\pm}$ are split by $2\Pi_\perp$ due to the presence of the effective field.

Interestingly, the spin is far more susceptible to transverse effective fields when the electric field dominates since $d_\perp \approx 50 d_\|$~\cite{VanOort1990}. In contrast, all spin-strain susceptibilities are comparable~\cite{Barson2017,Barfuss2019}. As a result, the average effect of a randomly oriented electric field leads to a large splitting with negligible common-mode shift (i.e. $\Pi_\perp \gg \Pi_\|$, on average), while in the case of strain the splitting is accompanied with a common-mode shift in the same order of magnitude (i.e. $\Pi_\perp \approx \Pi_\|$, on average). This key difference allows high-resolution spectroscopy of NV spins to differentiate between the electric field and strain contributions of the effective field, as we will show in the following.

Note that although we focus here on the case $B=0$ and $m_I = 0$, similar statements hold for the case $\gamma_{\rm NV} B_\| = \pm |A_{\rm HF}|$ and $m_I = \pm 1$ as well. 

\subsection{Magnetic dipole transition strengths}

In addition to shifting and splitting the hyperfine states, the effective field $\vb*{\Pi}$ also influences the dipole moment of the spin transitions. Experimentally, these spin transitions are probed by a MW field $\vb*{B}^{\rm{MW}}(t) =\vb*{B}^{\rm MW} \cos(\omega_{\rm MW} t)$ with frequency $\omega_{\rm MW}$ and (complex) amplitude $\vb*{B}^{\rm MW}$, which interacts with the corresponding magnetic dipole moment. As the dipole moment determines the polarization response of the transition, it is ultimately linked to the observed transition strengths.

Taking the common steps to transform into a rotating frame and applying the rotating wave approximation, the Rabi frequencies associated with the resonant spin transitions between $\ket{0}$ and $\ket{\pm}$ induced by the magnetic dipole interaction read
\begin{align}
\Omega_{0,\pm} = \frac{2\pi}{h} \qty|\mel**{\pm}{-\vb*{B}^{\rm{MW}} \cdot \vb*{\mu}}{0}| = \frac{2\pi}{h} \qty|\vb*{B}^{\rm{MW}} \cdot \vb*{\mu}_{0,\pm}| \ ,
\label{fml.matrixelement}
\end{align}
where the magnetic dipole moment operator is $\vb*{\mu} = -h \gamma_{\rm NV} \vb*{S} = - 2 \mu_B \vb*{S}$ with $\mu_B$ being the Bohr magneton. In equation\,\eqref{fml.matrixelement} we introduced the magnetic dipole matrix elements
\begin{align}
\vb*{\mu}_{0,\pm} = \mel**{\pm}{\vb*{\vb*{\mu}}}{0} = - 2 \mu_B \mel**{\pm}{\vb*{S}}{0}.
\label{fml.dipolemoment}
\end{align}
Evaluating this, we find  
\begin{align}
\begin{split}
\vb*{\mu}_{0,+} &= -2 \mu_B \qty[ \sin\qty(\varphi_\Pi/2), \cos\qty(\varphi_\Pi/2), 0] \ , \\
\vb*{\mu}_{0,-} &= -2 \mu_B \qty[ \cos\qty(\varphi_\Pi/2), -\sin\qty(\varphi_\Pi/2), 0] \ ,
\end{split}
\end{align}
showing that the azimuthal angle of the effective field is directly linked to both dipole moments. Note that the dipole moments are completely real, implying a linearly polarized response of the transitions. In contrast, when $B_\| \gg \Pi_\perp$, the eigenstates of Hamiltonian\,\eqref{fml.hamiltonian} are $\ket{0}, \ket{-1}$ and $\ket{+1}$, and the transitions show the familiar circularly polarized response.

According to equation\,\eqref{fml.matrixelement}, the relative orientation of the dipole moment to the MW field determines the Rabi frequency of the transition and therefore the observed ODMR response~\cite{Dreau2011}.
The observed transition strengths are then given by $\mathcal{A}_{0,\pm} \sim \Omega_{0,\pm}^2$. Writing the linearly polarized MW field amplitude as $\vb*{B}^{\rm{MW}} = [B_\perp^{\rm{MW}} \cos \varphi_{\rm{MW}}, B_\perp^{\rm{MW}} \sin \varphi_{\rm{MW}}, B_\|^{\rm{MW}}]$, where $\varphi_{\rm{MW}}$ is the azimuthal angle in the $xy$-plane (see figure\,\ref{fig1}(a)), the transition strengths between $\ket{0}$ and $\ket{\pm}$ read
\begin{align}
\mathcal{A}_{0,\pm} \sim \frac{(2\pi)^2}{h^2} \frac{(2 \mu_B B_\perp^{\rm MW})^2}{2} \qty( 1 \mp \cos(2\varphi_{\rm{MW}} + \varphi_\Pi)) \ .
\label{fml.amplitude}
\end{align}

Interestingly, the transition strengths contain information about the relative azimuthal angles of the effective field and the linearly polarized MW driving field.
Adapting the formalism of~\cite{Mittiga2018} this information can be derived from the transition imbalance
\begin{align}
\mathcal{I} = \frac{\mathcal{A}_{0,+} - \mathcal{A}_{0,-}}{\mathcal{A}_{0,+} + \mathcal{A}_{0,-}} = - \cos(2\varphi_{\rm{MW}} + \varphi_\Pi) \ .
\label{fml.imbalance}
\end{align}
Given that the azimuthal angle $\varphi_\Pi$ of the effective field $\vb*{\Pi}$ can be determined up to a reflection symmetry with respect to $2\varphi_{\rm MW}$ from the transition imbalance and the magnitude of $\Pi_\perp$ from the ODMR transition frequencies, we have established that both direction and magnitude of the transverse effective field can be extracted from high-resolution ODMR.

As mentioned earlier, the $m_I = \pm 1$ hyperfine projections are coupled by the effective field $\vb*{\Pi}$ as well. At $B=0$, however, the states with the same $m_I$ are split by $2|A_{\rm HF}|$, so that the transition imbalance due to the effective field is mostly suppressed. As we typically find $\Pi_\perp \ll |A_{\rm HF}|$ in our samples, we show in appendix\,\ref{sec.app1} that an approximate expression for the imbalance given by
\begin{align}
\mathcal{J} \approx - \frac{\Pi_\perp}{|A_{\rm HF}|} \cos(2\varphi_{\rm{MW}} + \varphi_\Pi) \ .
\label{fml.imbalance2}
\end{align}

\section{High-resolution spectroscopy}

\subsection{Experimental details}

In order to apply our findings above to investigate the effective field $\vb*{\Pi}$ and to determine the respective weights of the electric field and strain, we perform high-resolution spectroscopy on single NVs centers. For our experiments we use an electronic grade diamond (Element Six) implanted with $^{14}$N (dose \SI{e9}{ions/cm^2}, energy \SI{12}{keV} corresponding to an NV depth of $\sim$\,\SI{25}{nm}) and subsequently annealed using a high-temperature annealing process~\cite{Chu2014}. Note that several fabrication steps, e.g. etching, were performed on the sample~\cite{Teissier2014}, possibly causing a larger intrinsic strain in the diamond compared to an untreated sample.

To perform low-power ODMR measurements we utilize established techniques for optical initialization and readout of the NV spin state using a home-built confocal microscope setup~\cite{Teissier2014, Barfuss2015}. Three pairs of magnetic coils allow us to apply an external magnetic field with full vector control (see appendix\,\ref{sec.app2} and~\cite{Barfuss2017}). Manipulation of the NV spin in the ground state is realized by MW magnetic fields. More precisely the circularly polarized component of the MW field projection transverse to the NV axis ($B^{\rm{MW}}_\perp$, see figure\,\ref{fig1}(a)) allows us to drive transitions between different spin levels with the same $m_I$, as illustrated with the red arrow in figure\,\ref{fig1}(b). We realize this MW driving by applying an AC current to a gold wire with a diameter of $\sim$\,\SI{30}{\micro m} close to the NVs under study (see figure\,\ref{fig1}(c)), thereby coupling the NVs to the near-field of the MW source. This configuration leads to an approximately linearly polarized MW field at the NVs' locations.

Note that geometric considerations for our setup allow us to determine the MW polarization angle required to interpret the experimental data (compare to equation\,\eqref{fml.imbalance}). We find $\varphi_{\rm MW} \approx \SI{90}{\degree}$ for the specific orientation of a particular single NV ('NV1'). However, we will later present a technique to determine this MW polarization angle without the requirement of such a geometric consideration. This technique, which is based on the controlled rotation of a large transverse magnetic field, yields very good agreement with our a priori determination of $\varphi_{\rm MW}$.

\begin{figure*}[htb]
	\centering
		\includegraphics{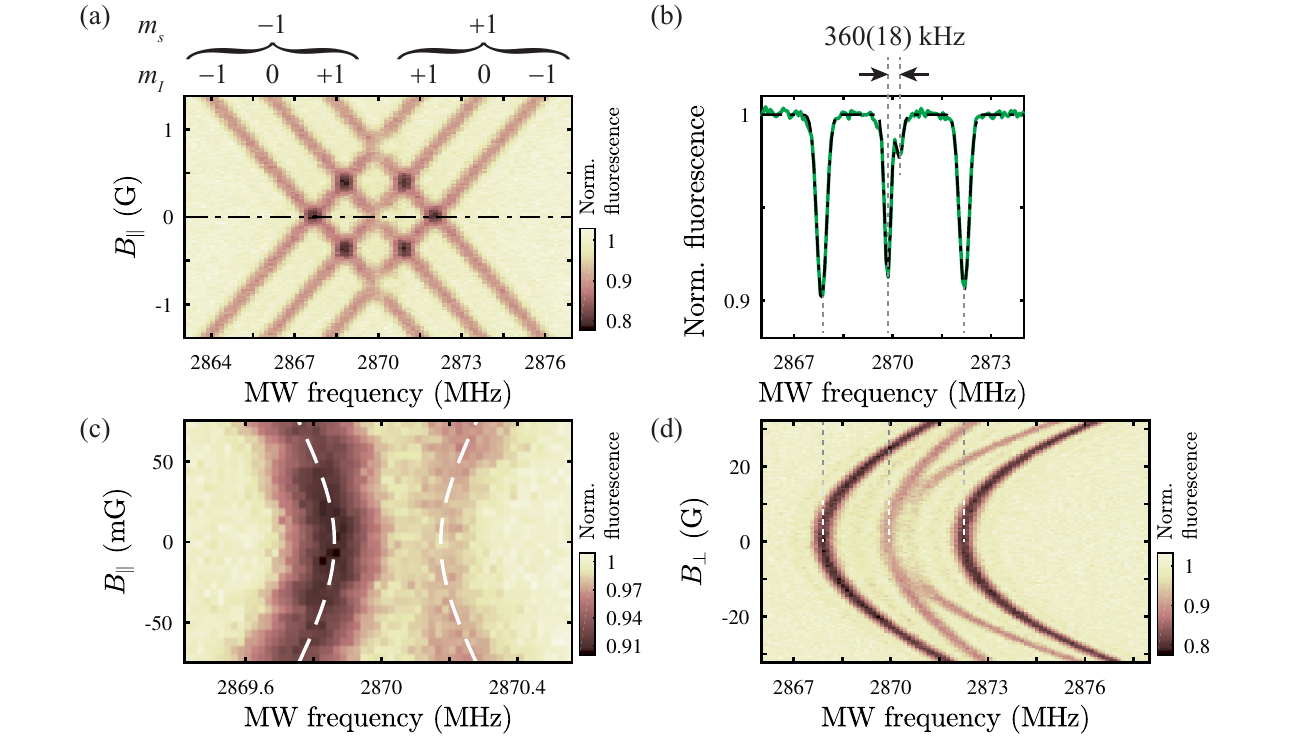}
	\caption{Influence of the external fields on the electron spin transitions.
	(a) A magnetic field $B_\|$ parallel to the NV axis results in a linear Zeeman splitting of the six possible spin transitions from $\ket{0}$ to $\ket{\pm 1}$. However, when two transitions with the same nuclear spin projection $m_I$ cross, we observe a reduced ODMR contrast, indicating a coupling of the corresponding states.
	(b) Line cut for $B_\| = \SI{0}{G}$ with reduced MW power including a Gaussian fit to the data (black). The central line is split by \SI{360(18)}{kHz} into two transitions with an imbalance of $\mathcal{I} = \SI{-58(5)}{\%}$.
	(c) Zoom into the $m_I = 0$ transitions around $B_\| \approx \SI{0}{G}$ visualizing a clear level anti-crossing, indicative of a coupling between both hyperfine levels. 
	(d) Applying a transverse magnetic field $B_\perp$ while maintaining $B_\| \approx \SI{0}{G}$ mixes the spin levels and results in a second order energy shift (see section\,\ref{sec.MWdetermination}).}
	\label{fig2}
\end{figure*}

\subsection{Spectroscopy around zero magnetic field}

To characterize our sample we first investigate the response of NV1 (inhomogeneous dephasing time of $T_2^* \sim\,\SI{2}{\micro s}$) to external magnetic fields. Starting with a magnetic field parallel to the NV axis, we record pulsed ODMR measurements~\cite{Dreau2011} for various values of $B_\|$ (see figure\,\ref{fig2}(a)). The resulting spectrum shows six hyperfine resolved spin transitions. Due to the Zeeman effect, the three nuclear spin projections with $m_s = +1$ ($m_s = -1$) show a positive (negative) dispersion with magnetic field, i.e. shift to larger (smaller) frequencies. The order of the nuclear spin projections can be established from Hamiltonian\,\eqref{fml.hamiltonian} (see figure\,\ref{fig2}(a)). At and above the maximum values of $B_\|$ we apply, all six ODMR transitions show the same contrast, as the nuclear spin states are equally (thermally) populated, and the $m_s = -1$ ($m_s = +1$) transitions have right (left) circularly polarized response, whereas we apply a linearly polarized MW field (an equal superposition of right and left circular polarization). At the transition crossings where states have different values of $m_I$, the states do not mix and the transition strengths, which are related to the contrast of both transitions, sum together, resulting in twice the fluorescence drop compared to a single transition. When two crossing states have the same nuclear spin projection $m_I$, however, the states do mix and we observe stark differences to the case of states of unequal $m_I$ crossing.

To investigate this observation in more detail and to verify the coupling of the corresponding states, we record a high-resolution ODMR line cut at $B = 0$ (see figure\,\ref{fig2}(b)). The spectrum shows a clear splitting of the central transition into two peaks, both having different contrast. This effect is attributed to the influence of the effective field $\vb*{\Pi}$, which mixes the $\ket{\pm 1}$ spin states, as explained previously. Fitting the experimental data with a superposition of four Gaussian functions allows us to extract a splitting of $2\Pi_\perp = \SI{360(18)}{kHz}$ and a transition imbalance of $\mathcal{I} = \SI{-58(5)}{\%}$. Note that for consistency, we also obtained the same results by using continuous-wave, instead of a pulsed ODMR (data not shown).

To ensure that our findings are not masked by any residual parallel magnetic field, we perform high-resolution ODMR measurements with the parallel magnetic field component in the vicinity of the $m_I = 0$ transition crossing (see figure\,\ref{fig2}(c)). We clearly resolve an anti-crossing of the two transitions (illustrated by the dashed white line), consistent with the discussed coupling of the corresponding hyperfine states. Note that all the measurements presented in figure\,\ref{fig2} and the following are recorded separately and hence under slightly different experimental conditions (e.g. temperature), leading to small variations in the zero-field splitting~\cite{Acosta2010}.

\begin{figure*}[htb]
	\centering
		\includegraphics{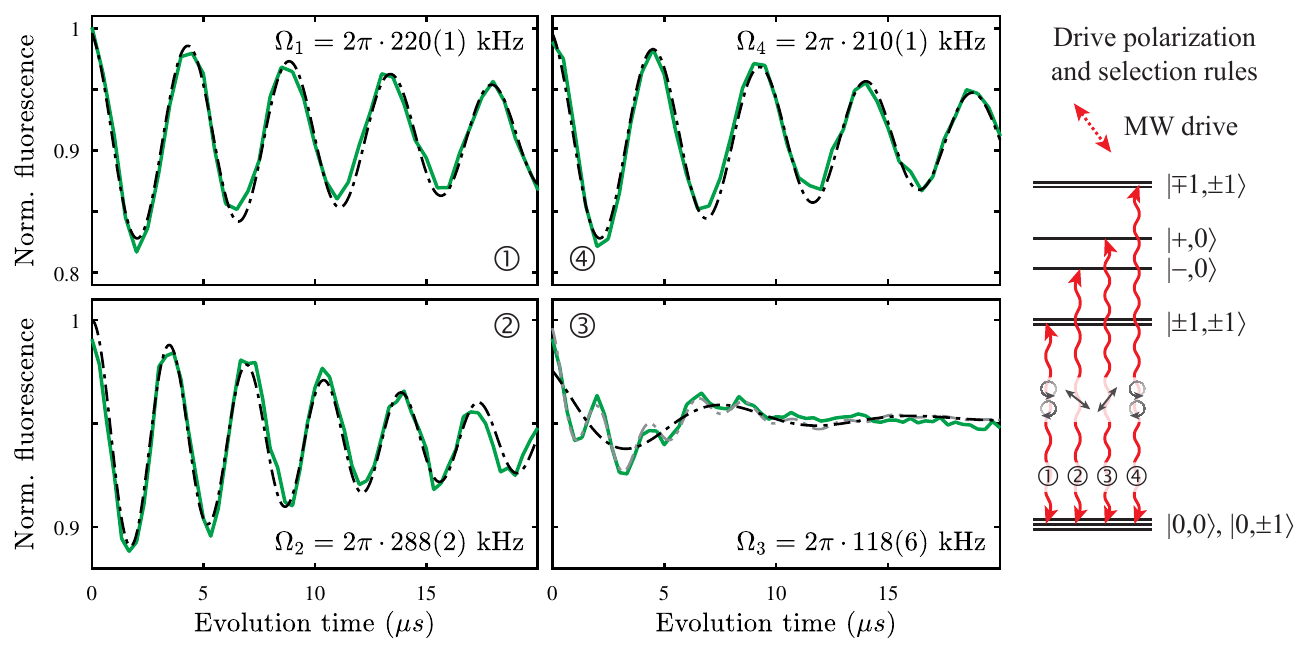}
	\caption{Comparison of the transition strengths for $B_\| = \SI{0}{G}$. Driving Rabi oscillations on each of the four transitions shown on the right under the same experimental conditions allows us to directly compare the transition strengths. To a good approximation we find $\Omega_{1}^2 = \Omega_{4}^2 = \tfrac{1}{2}(\Omega_{2}^2 + \Omega_{3}^2)$, as expected (see text). Note that panel~\ding{174} contains two oscillation frequencies, the slowly oscillating component of transition~\ding{174} (black) and a higher frequency caused by transition~\ding{173} (see text). The level diagram additionally shows the polarization of the MW drive and the polarization response of each transition.}
	\label{fig3}
\end{figure*}

\subsection{Comparison of transition strengths}

To quantify the strengths of the involved ODMR transitions, we conduct Rabi oscillation experiments on each transition using the same MW and laser power (see figure\,\ref{fig3}). Equation\,\eqref{fml.amplitude} directly yields a relation between the Rabi frequencies $\Omega_i$ ($i=1,2,3,4$) of the four different transitions as
\begin{align}
\Omega_1^2 = \Omega_4^2 = \frac{1}{2} \qty( \Omega_2^2 + \Omega_3^2 ) \ .
\end{align}
Here, we used the labels for the transitions according to figure\,\ref{fig3} and defined $\Omega_1^2 = \Omega_4^2 = (2\pi)^2 \cdot 2 (\mu_B B_\perp^{\rm{MW}})^2/h^2$. Our experimental findings agree well with this prediction, as $\Omega_{1} = 2\pi \cdot \SI{220(1)}{kHz} \approx \Omega_{4} = 2\pi \cdot \SI{210(1)}{kHz}$ and $(\frac{1}{2} ( \Omega_{2}^2 + \Omega_{3}^2))^{1/2} = 2\pi \cdot \SI{220(3)}{kHz}$, where $\Omega_2 = 2\pi \cdot \SI{288(2)}{kHz}$ and $\Omega_3 = 2\pi \cdot \SI{118(6)}{kHz}$. 

The measured transition strengths are directly linked to the polarization response of each transition (see figure\,\ref{fig3}, right). Transitions~\ding{172} and~\ding{175} each involve two transitions of $\Delta m_s = \pm 1$, such that they comprise both circular polarization responses. In contrast, the transitions~\ding{173} and~\ding{174} correspond to transition from $\ket{0}$ to the eigenstates\,\eqref{fml.eigenstates}, which are superpositions of $\ket{\pm 1}$ and yield a linearly polarized response of the transitions. Since the polarization of the MW drive is approximately linear, the overlap of the drive polarization and transition dipole determine the transition strengths (refer to equation\,\eqref{fml.matrixelement}).

We note that in the experimental data for transition~\ding{174}, there are two oscillation frequencies present (see gray fit). The slowly oscillating component highlighted by the black line corresponds to driving of transition~\ding{174}, while the quickly oscillation component originates from off-resonant driving of transition~\ding{173} which is neglected in the black line.

Additionally, the slight mismatch between $\Omega_1$ and $\Omega_4$ is a direct consequence of the coupling between the $m_s = \pm 1$ spin states with the same nuclear spin projection ($m_I = \pm 1$) present at $B=0$. This discrepancy is caused by the transition imbalance described in equation\,\eqref{fml.imbalance2}. This slight mixing means that the transitions show an elliptically polarized response, rather than a purely circularly polarized response.

\begin{figure*}[htb]
	\centering
		\includegraphics{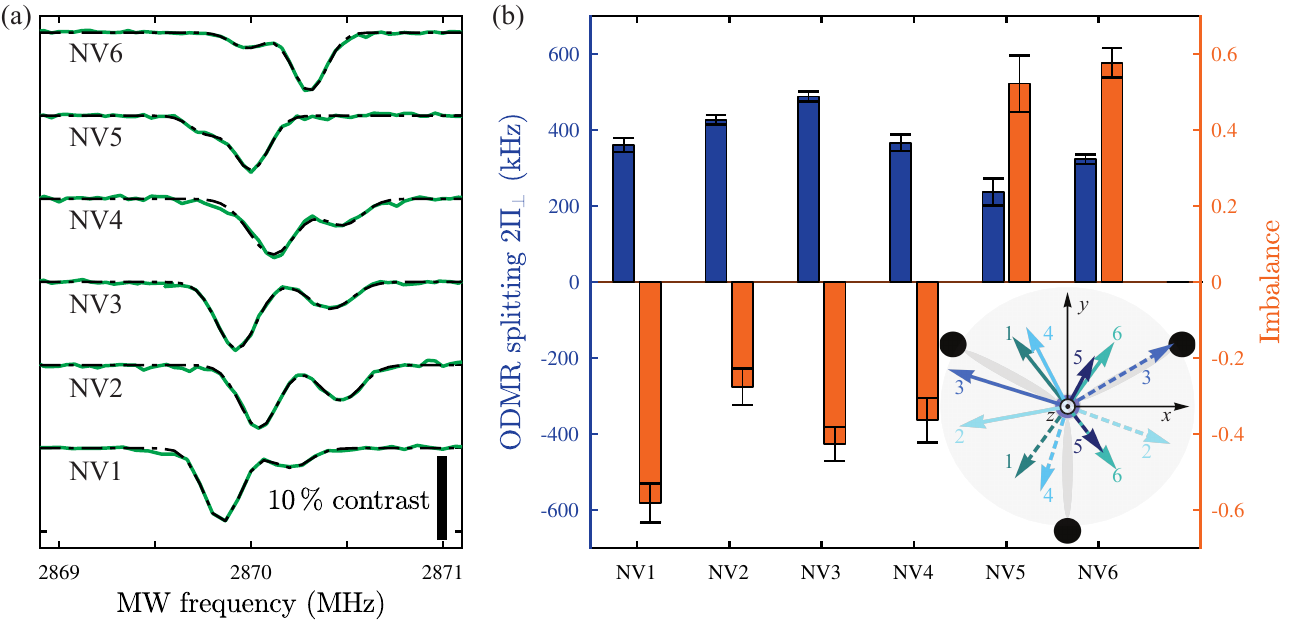}
	\caption{Characterization of the effective fields of selected single NV centers.
	(a) High resolution ODMR spectra for $B_\| = \SI{0}{G}$ for several single NV centers showing a splitting of the $m_I = 0$ levels and transition imbalances of the corresponding transitions.
	(b) Summary of the extracted splittings and transition imbalances from (a) and schematic illustration of the transverse effective field component for each NV (numbers refer to NV labels). Each dataset is consistent with two possible values for $\varphi_\Pi$ and both possible effective field orientations are shown for each NV.}
	\label{fig5}
\end{figure*}

\subsection{Characterization of individual NV centers}

Having established our technique to probe and characterize the effective field $\vb*{\Pi}$ of a single NV, we now apply this method to investigate the field environment of a selection of individual NVs. For that we perform high-resolution ODMR measurements at $B=0$ as outlined earlier and studied the splitting and transition imbalance of the $m_I = 0$ transitions (see figure\,\ref{fig5}(a)). We find that the splitting and transition imbalance is different for each NV under study, indicating a different effective field environment for each defect. Comparing the mean transition frequency for each NV with the averaged transition frequency over all NVs, we find that the observed shifts and splittings are of the same order of magnitude, which indicates that strain is the dominant contribution to the effective field. This observation differs from the findings of~\cite{Mittiga2018}, where the electric field originating from charge impurities was identified as the main effective field contribution in their samples. However, our method does not allow a complete determination of the effective field, since we are not able to determine the parallel component $\Pi_\|$. The zero-field splitting depends sensitively on temperature~\cite{Acosta2010}, i.e. both temperature and the parallel effective field component have the same effect on the observed transition frequencies. Nevertheless, we can identify strain as the major contribution to the effective field, as we typically observe peak-to-peak temperature fluctuations of \SI{0.3}{K} in our setup corresponding to temperature-induced shifts of $\sim$\,\SI{23}{kHz}, much smaller than the ones observed in figure\,\ref{fig5}.

Using Gaussian fits we extract the splittings $2\Pi_\perp$ and transition imbalances for each ODMR spectrum in figure\,\ref{fig5}(a) and summarize the results in figure\,\ref{fig5}(b). Using equation\,\eqref{fml.imbalance} allows us to visualize the transverse effective field components in the NV frames up to a reflection symmetry with respect to $2\varphi_{\rm MW}$ as presented in the inset of figure\,\ref{fig5}(b). Here, the transverse effective field amplitudes are normalized with respect to NV1 and we used the MW polarization angle of NV1 as determined in the following section\,\ref{sec.MWdetermination}.
Note that figure\,\ref{fig5} shows NVs with all four possible orientations in the diamond lattice, which experience different relative MW polarization angles accordingly and therefore show a different symmetry behavior. Once we know the polarization angle for a single NV, geometric considerations allow us to infer the corresponding angles for the other NV orientations. To determine $\varphi_\Pi$ without ambiguity, one can e.g. conduct a second set of measurements with a different MW polarization angle $\varphi_{\rm MW}$.

\begin{figure*}[htb]
	\centering
		\includegraphics{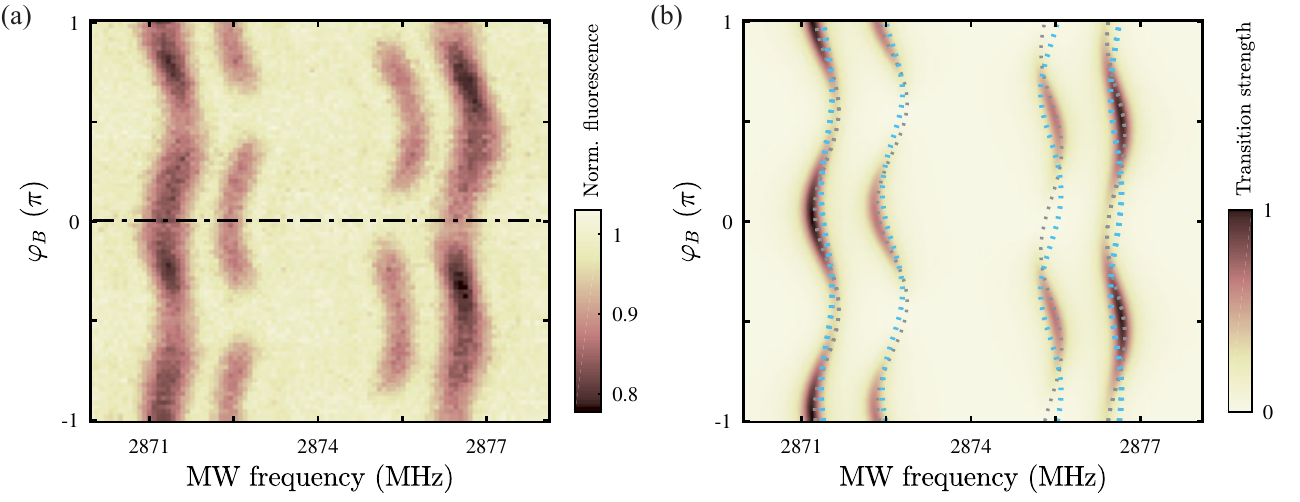}
	\caption{Determination of the MW polarization angle $\varphi_{\mw}$ by rotating the transverse magnetic field component around the quantization axis of NV1.
	(a) Experimental data for $B_\perp = \SI{32}{G}$ and $B_\| = \SI{0}{G}$. The transition strengths of all states oscillate. Periodic 'wiggles' in the transition frequencies are caused by intrinsic strain in the diamond lattice and imperfect rotation of the magnetic field (see text).
	(b) Comparing the experimental data with the calculation based on Hamiltonian\,\eqref{fml.hamiltonian} allows for determining the MW polarization angle in the transverse plane in the reference frame of NV1 to be $\varphi_{\mw} = \SI{89.2(8)}{\degree}$ with respect to the $x$-axis. To quantify the relative contributions of intrinsic strain and the elliptic magnetic field rotation we additionally show the simulated transition frequencies for considering both aspects (gray dotted) and for only considering strain while the magnetic field rotation is assumed to be perfectly circular (light blued dotted).}
	\label{fig4}
\end{figure*}

\section{\label{sec.MWdetermination}Determining the MW polarization angle}

We finalize our spectroscopy study of the effective field by establishing a method to experimentally determine the MW polarization angle. To do so, we exploit the response of the hyperfine spin transitions to a purely transverse magnetic field. Because a transverse magnetic field couples $\ket{\pm 1}$ states in second order, the transition frequencies show a quadratic dependence on the field (see figure\,\ref{fig2}(d) for data on NV1). Similarly to figure\,\ref{fig2}(a), the spectrum shows in total six transitions. However, the corresponding states of the two outer transitions are degenerate, resulting in four resolvable transitions with the outer two having twice the contrast relative to the inner ones. The inner transitions approach the dominant outer transitions for large fields, indicating a tilting of the spin quantization axis towards the transverse field axis.

At $B_\perp = \SI{32}{G}$ (the largest transverse magnetic field amplitude achievable in our setup) we study the influence of the azimuthal angle of the transverse field $\varphi_B$ while setting $B_\| = \SI{0}{G}$ (within our experimental resolution). In this situation, we change $\varphi_B$ and perform pulsed ODMR measurements on the hyperfine transitions (see figure\,\ref{fig4}(a)). We observe clear oscillations of the contrasts of the four transitions as a function of $\varphi_B$. Specifically, the contrasts of the two lower frequency transitions  (at $\sim$\,\SI{2871}{MHz} and $\sim$\,\SI{2872.5}{MHz}) oscillate in phase, as do the contrasts of the two higher frequency transitions (at $\sim$\,\SI{2875.5}{MHz} and $\sim$\,\SI{2877}{MHz}). However, the contrast-oscillations of these two pairs of transitions oscillate out of phase as a function of  $\varphi_B$, with a phase shift $\sim$$\,\pi/2$, i.e. the lower frequency transitions have highest contrast when the higher frequency transitions have low contrast, and vice versa.

Using a similar theoretical model as previously introduced, we calculate the transition strengths based on Hamiltonian\,\eqref{fml.hamiltonian}. As we consider the regime $\gamma_{\rm NV} B_\perp \gg \Pi_\perp, |A_{\rm HF}|$, the effective field and hyperfine coupling can be neglected in first order. The eigenstates of the Hamiltonian thus read $\ket{\tilde{0}} \approx \ket{0}$ and $\ket{\tilde{\pm}}$, where $\ket{\tilde{\pm}}$ are mixed states of $\ket{\pm1}$ due to the presence of the transverse magnetic field. In analogy to our earlier results, we find the transition strengths
\begin{align}
\mathcal{A}_{0,\tilde{\pm}} \sim \Omega_{0,\tilde{\pm}}^2 = \frac{2\pi}{h} \qty| \vb*{B}^{\rm{MW}} \cdot \vb*{\mu}_{0,\tilde{\pm}} |^2 \simeq \frac{(2\pi)^2}{h^2} \frac{(2\mu_B B_\perp^{\rm MW})^2}{2} \qty( 1 \pm \cos(2\varphi_B - 2\varphi_{\rm{MW}})) \ ,
\label{fml.mwpolarization}
\end{align}
where we have used the magnetic dipole matrix elements $\vb*{\mu}_{0,\tilde{\pm}} = -2\mu_B \mel{\tilde{\pm}}{\vb*{S}}{0}$. According to equation\,\eqref{fml.mwpolarization} rotating the transverse magnetic field component by changing $\varphi_B$ leads to oscillations in the transition strengths of the involved transitions. The $\pi$-periodicity of these oscillations is induced by the periodic change of the overlap of the MW polarization and the dipole moments, and is mirrored by the alternating ODMR contrast in the experiment. The phase offset of these oscillations is related to the MW polarization angle $\varphi_{\rm MW}$. Thus, by analyzing the experimental data we find $\varphi_{\rm{MW}} = \SI{89.2(8)}{\degree}$ in the reference frame of NV1.

Equation\,\eqref{fml.mwpolarization} reproduces the experimentally observed oscillations in the transition intensity. According to the presented model, however, the frequencies of the observed transitions should not be affected by varying $\varphi_B$. Instead, the observed 'wiggles' in the transition frequencies are attributed to the influence of the neglected effective field and the effect of an imperfect, elliptical rotation of the transverse magnetic field amplitude. We numerically modeled both aspects by simulating our experiment based on Hamiltonian\,\eqref{fml.hamiltonian} (see figure\,\ref{fig4}(b)). For the simulation, we used the effective field parameters extracted from figure\,\ref{fig2}(b) and a realistic ellipticity of the transverse magnetic field rotation characterized by a flattening of $f = 0.039$. With that we are able to reproduce the wiggling of the transition frequencies and the asymmetric shift of the transition contrast. To characterize the relative contribution of both effects, figure\,\ref{fig4}(b) further shows the simulated transition frequencies. These can be compared to the calculated transition frequencies for the case where only the known effective field is considered while the magnetic field rotation is assumed to be perfectly circular. From this comparison it becomes apparent that only the pair of higher frequency transitions is significantly affected by the ellipticity of the magnetic field, as the states involved in these transitions are more susceptible to magnetic fields than the ones pertaining to the low-frequency pair of transitions. Moreover, the amplitude of the wiggles induced by the ellipticity is very similar to the amplitude of the wiggles induced by the effective field, thus both effects are of comparable order. Nevertheless, we conclude from our simulations, that this experimental imperfection does not affect any of the other findings we report on here.

\section{Conclusion}

In this paper, we presented high-resolution, low-power ODMR spectroscopy studies on single NV defect centers in diamond to characterize their local effective field environment. Our approach is based on a detailed examination of the NV spin's allowed magnetic dipole transitions, which are affected by the interaction of the MW probing field and the intrinsic effective field. Comparing with previous studies on NV ensembles in more strongly doped diamonds, we found that in our case of single NVs in high purity diamonds, strain is the major contribution to the effective field. In addition, we demonstrated a new method for performing single spin-based, linear polarization analysis of MW fields based on low-field, high-resolution ODMR in well-controlled bias fields.

The fact that our conclusions on the nature of the local effective field of the single NV spins we investigated differ from recent studies on NV ensembles~\cite{Mittiga2018}, highlights the importance of characterizing such fields in a quantitative and effective way for future quantum technology development. Such characterization is then particularly relevant for NV-based quantum sensing applications where low-field operation is key. Examples for these include nanoscale magnetic imaging of magnetically sensitive samples~\cite{Tetienne2014a}, or NV-based low-field techniques like zero and ultra-low field NMR~\cite{Jiang2019a}.
The novel MW polarization analysis we demonstrated could find applications in NV-based MW imaging~\cite{Appel2015,Horsley2018a}, which until now was only demonstrated for sensing of circularly polarized MWs~\cite{Wang2015}. Our results extend these capabilities and the existing toolset of NV-based quantum sensing modalities and would in principle allow for determining the full polarization state of MW fields with nanoscale resolutions, which has relevance in MW electronics~\cite{Rosner2002} or spintronics devices~\cite{Andrich2017a}.

\begin{acknowledgments}
We thank N. Yao, D. Budker, B. Kobrin, S. Hsieh and J. Wood for fruitful discussions and valuable input. We gratefully acknowledge financial support through the NCCR QSIT (Grant No. 185902), through the Swiss Nanoscience Institute, through  the EU Quantum Flagship project ASTERIQS (Grant No. 820394) and through the Swiss NSF Project Grant No. 169321.
\end{acknowledgments}

\appendix
\section{\label{sec.app1}Derivation of transition imbalances for $m_I = \pm 1$}

In order to derive the transition imbalance for the $m_I = \pm 1$ hyperfine projections at $B=0$~as stated in equation\,\eqref{fml.imbalance2}, we consider the Hamiltonian
\begin{align}
\mathcal{H}/h = \qty(D_0 + \Pi_z) S_z^2 \mp |A_{\rm{HF}}| S_z + \Pi_x \qty(S_y^2-S_x^2) + \Pi_y \qty(S_x S_y + S_y S_x) \ .
\end{align}
Note that the sign of the hyperfine interaction is flipped compared to equation\,\eqref{fml.imbalance2} as $A_{\rm HF} < 0$. Following~\cite{Mittiga2018}, we use the same procedure as in the main text and first calculate the corresponding eigenstates
\begin{align}
\begin{split}
\ket{-} &= \frac{1}{\sqrt{1+\lambda^2}} \qty( e^{i \varphi_\Pi} \ket{+1} + \lambda \ket{-1} ) \ , \\
\ket{+} &= \frac{1}{\sqrt{1+\lambda^2}} \qty( \lambda e^{i \varphi_\Pi} \ket{+1} - \ket{-1} ) \ ,
\end{split}
\end{align}
where we have defined
\begin{align}
\lambda = \frac{|A_{\rm HF}|}{\Pi_\perp} \qty( \sqrt{1 + \qty( \frac{\Pi_\perp}{|A_{\rm HF}|})^2} - 1) \ .
\end{align}
Using equations\,\eqref{fml.matrixelement} and\,\eqref{fml.dipolemoment} we find for the transition strengths
\begin{align}
\mathcal{A}_{0,\pm} \sim \frac{(2\pi)^2}{h^2} \frac{(2\mu_B B_\perp^{\rm MW})^2}{2} \frac{1+\lambda^2\mp 2\lambda \cos(2\varphi_{\rm MW} + \varphi_\Pi)}{2(1+\lambda^2)} \ . 
\end{align}
Thus, the transition imbalance for the $m_I = \pm 1$ nuclear spin projections $\mathcal{J} = \frac{\mathcal{A}_{0,+} - \mathcal{A}_{0,-}}{\mathcal{A}_{0,+}+\mathcal{A}_{0,-}}$ is given by
\begin{align}
\mathcal{J} = - \frac{2 \lambda}{1+\lambda^2} \cos(2\varphi_{\rm MW} + \varphi_\Pi) \ .
\end{align}
Considering the case $\Pi_\perp \ll |A_{\rm HF}|$, we can approximate $\lambda \approx \frac{\Pi_\perp}{2 |A_{\rm HF}|}$ and then find to first order that the imbalance is
\begin{align}
\mathcal{J} \approx - \frac{\Pi_\perp}{|A_{\rm HF}|} \cos(2\varphi_{\rm MW} + \varphi_\Pi) \ ,
\end{align}
as stated in equation\,\eqref{fml.imbalance2} in the main text.

\section{\label{sec.app2}Alignment and control of the magnetic field}

Our experimental setups comprises three pairs of coils arranged in a Helmholtz-like setup, i.e. the spatial separation between corresponding coils matches their diameter ($X$, $Y$-pairs) or their radius ($Z$-pair)~\cite{Barfuss2017}. Each pair is driven by a constant-current source (Agilent E3644A) enabling software-based three-dimensional magnetic field control. We performed the calibration of the magnetic field calibration with a Teslameter (Projekt Elektronic, FM 302 with transverse probe AS-NTM).

To align the magnetic field to a desired NV orientation we used the procedure described in detail in the supplementary material of~\cite{Thiel2016}. This method relies on the controlled rotation of the magnetic field in space and the resulting effect on the NV's ODMR frequency. Comparing the experimental data with our simulations allows us to estimate two important parameters: First, the simulations show that the achieved alignment uncertainty is within $<\ang{0.2}$ to a desired direction (otherwise the outer two degenerate states would be split in figure\,\ref{fig5}(a)). Second, we estimated that the ellipticity of the magnetic field rotation is characterized by a flattening $f < 0.04$, otherwise the observed wiggles in figure\,\ref{fig5}(a) would be larger. The fact that we observed such an ellipticity may be attributed to uncertainty in the calibration of the coils.

Both facts, however, only appear in the data of figure\,\ref{fig5}, as we applied and rotated a large perpendicular magnetic field of $B_\perp = \SI{32}{G}$ in this case. For the other measurements that we used to extract the effective field parameters, the magnetic fields are static and only used to cancel external magnetic fields (i.e. earth magnetic field). Thus, the extracted effective field parameters are not affected.

\bibliography{NVHighResolutionSpectroscopy_2}
\bibliographystyle{apsrev4-1}

\end{document}